\documentclass[format=acmsmall, review=false, screen=true, ]{acmart}

\usepackage{booktabs} 

\setcopyright{rightsretained}
\usepackage{url}  

\usepackage{graphicx}  

\usepackage{booktabs} 
\usepackage{xspace}
\usepackage{color}

\usepackage{subcaption}

\usepackage[autolanguage,np]{numprint}
\usepackage{color}
\usepackage{flushend}
\usepackage{amssymb}
\usepackage{xspace}
\usepackage[english]{babel}
\usepackage[utf8]{inputenc}

\usepackage{paralist}
\usepackage{amsmath}
\usepackage{algorithm}
\usepackage[noend]{algpseudocode}
\algloopdefx[loop]{Iterate}
[1][]{{\bf iterate} }

\newcommand{\evidense}{\textsc{EviDense}\xspace}

\newcommand{\mabed}{\textsc{MABED}\xspace}
\newcommand{\edcow}{\textsc{EDCoW}\xspace}

\newtheorem{definition}{Definition}
\begin{document}


\title{\textsc{EviDense}: a Graph-based Method for Finding Unique High-impact Events with Succinct Keyword-based Descriptions }

\author{Oana Balalau}
\affiliation{%
\institution{Inria and École Polytechnique}
\country{France}}
\email{oana.balalau@inria.fr}
\author{Carlos Castillo}
\affiliation{%
\institution{Universitat Pompeu Fabra}
\country{Spain}}
\email{carlos.castillo@upf.edu}
\author{Mauro Sozio}
\affiliation{
\institution{T\'el\'ecom Paris University}
\country{France}}
\email{sozio@telecom-paristech.fr}


\begin{abstract}

Despite the significant efforts made by the research community in recent years, automatically acquiring valuable information about high impact-events from social media remains challenging.  We present \textsc{EviDense}, a graph-based approach for finding high-impact events (such as disaster events) in social media.
One of the challenges we address in our work is to provide for each event a succinct keyword-based description, containing the most relevant information about it, such as what happened, the location, as well as its timeframe. We evaluate our approach on a large collection of tweets posted over a period of 19 months, using a crowdsourcing platform. Our evaluation shows that our method outperforms state-of-the-art approaches for the same problem, in terms of having higher precision, lower number of duplicates, and presenting a keyword-based description that is succinct and informative.
We further improve the results of our algorithm by incorporating news from mainstream media. 

A preliminary version of this work was presented as a 4-pages short paper at ICWSM 2018~\cite{Balalau2018}.
\end{abstract}


\maketitle

\section{Introduction}

Social media have been playing increasingly a major role during crises and disasters. For example, the American Red Cross (ARC) pointed out the effectiveness of social media and mobile apps in handling emergency situations such as those generated by a disaster event (e.g. earthquakes, mass shootings, etc.)\footnote{http://www.redcross.org/news/press-release/More-Americans-Using-Mobile-Apps-in-Emergencies}.
Automatically acquiring valuable information about high-impact events from social media would be extremely valuable, however, it presents non-trivial challenges. Social content is often noisy, inconsistent and ambiguous. Tweets, for example, are short and written by non-experts using informal language while containing typos and abbreviations. Moreover, relevant information is often intertwined with noisy or non-interesting content such as spam or posts by so-called Twitter bots. Another additional difficulty when analyzing data from Twitter lies on the fact that not all tweets are publicly available, with only a small fraction of them being accessible through rate-limited APIs. Indeed, despite the significant efforts made by the research community --see, e.g., surveys ~\cite{Imran2014,FarzindarK15,castillo2016} -- automatically acquiring valuable information about high-impact events from social media remains challenging. 

One of the challenges we address in our work is how to provide a succinct keyword-based description of high-impact events containing the most relevant information about the events, such as what happened, where, and when. According to a survey by the US Congressional Service, the administrative cost for monitoring multiple social media sources, which typically produce large amounts of noisy data, is significant~\cite{lindsay2011social}. Therefore, in order to alleviate the burden of analyzing social content, a succinct and informative description of the events is needed.

Our approach consists of the following steps:
\begin{inparaenum}[i)]
\item filtering of the tweets by retaining only those containing at least one term in a given lexicon. In our work, we use the lexicon constructed in~\citet{Olteanu2014} which contains terms related to disaster events, however, any lexicon can be used; 
\item finding locations whose number of occurrences in tweets deviates significantly on a given time window from their expected frequency;
\item a graph mining approach for selecting the relevant keywords in the description, based on a novel definition of a clique and a quasi-clique in a weighted graph. 
\end{inparaenum}
To ensure that the description is succinct we enforce an upper bound (specified in input) on the number of keywords to be selected.

We call our approach \evidense, as our graph mining approach is based on finding ``dense'' regions in the graph representing the co-occurrence of keywords in the tweets. 
We evaluate this algorithm against state-of-the-art approaches on a collection of tweets covering the period between November 2015 and May 2017, by means of a crowdsourcing platform. Our evaluation represents one of the most extensive evaluations for an event detection algorithm in social media.
It shows that our approach outperforms the baselines both in terms of precision (at $k$) and number of duplicates, while the keyword-based description provided by our algorithm is succinct and informative. Moreover, we show how to further improve the results of our algorithm by incorporating news from mainstream media.
Given these results, we consider \evidense could represent a valuable tool for analyzing both social content and news articles from mainstream media, as well as for studying how they compare. It could also be used to boost the performance of other approaches, such as automatic classifiers. For this purpose, we released a collection of tweets containing mentions of the disaster events identified by our approach.

To summarize, the contributions of this paper are:
\begin{itemize}
\item We propose a completely unsupervised technique for finding  with high precision mentions of high-impact events.
\item We show that our approach can incorporate information from mainstream media in order to improve event detection in social media. 
\item We propose a new definition of a clique and quasi-clique in a weighted graph and we compare its performance to other dense subgraph definitions. The quasi-clique is a building block of our event detection algorithm and gives a succinct description of an event. 
\item We show that our approach is versatile and can be used for finding other mentions of high-impact events, such as political events.  
\item We provide a dataset containing tweets that were retrieved using the disasters' keyword descriptions computed by our algorithm. 
\end{itemize}

The rest of the paper is organized as follows:
In Section~\ref{sec:related} we outline the related work; in Section~\ref{sec:algo} we present our algorithm, while we compare it against state-of-the-art approaches in Section~\ref{sec:eval}. In Section~\ref{sec:dense}, we compare different definitions of dense subgraphs for the task of event description in social media. 
Finally, we recap and discuss some possible use cases in Section~\ref{sec:future}.

\section{Related Work}\label{sec:related}

\textbf{Event detection.}
Several approaches have been proposed for event detection in social media, which can be classified according to whether they focus on a particular pre-specified class of events (e.g., such as earthquakes, computer security breaks) or whether such a class is not specified. Another criteria that can be used to classify the related work is the main technique used. The most successful techniques that have been employed for this task include clustering-based methods, graph mining approaches, wavelet, and supervised learning methods based on SVMs or gradient boosted decision trees.
There are extensive surveys on this topic~\cite{FarzindarK15,Imran2014,Madani2014,Nurwidyantoro2013}.
Approaches that deal with a pre-specified class of events typically use classifiers in order to label tweets \cite{Sakaki,Imran2014} or aggregations of tweets \cite{Popescu2011,Ritter2015}. In~\cite{Imran2014}, the
authors present a platform for automatic classification of messages during a disaster
event. More precisely, first tweets are crawled using certain criteria like location or
keywords and then human annotators label tweets using different categories, such
as ``needs''. An automatic classifier is trained based on the labels and retrained as
new labels arrive. The authors emphasize that automatic classification using a pre-existing dataset is not a good solution as disaster events can have very specific aspects that differentiate them. In order to facilitate the task of finding mentions of disasters, we propose an unsupervised approach for finding such mentions that can scale to very large datasets. Therefore, in the following we focus on other unsupervised event detection methods that allow a large-scale analysis.  




\citet{WengL11} propose an event detection algorithm (\emph{EDCoW}) based on clustering of the wavelet-based signal of words. More precisely, the frequency of words in text over time is a time series, which can be processed using wavelet analysis. The authors filter trivial words using auto-correlation of signals and cluster the remaining words employing as a similarity metric the cross-correlation of signals. Clusters are finally ranked according to a score which takes into account the cross-correlation of words and the total number of words associated with each cluster. 

\citet{Cordeiro2012} proposes an algorithm based on continuous wavelet transformation and Latent Dirichlet Allocation (LDA), a widely-used unsupervised method for topic modeling.
In the first step, the frequency of hashtags over time is analyzed using continuous wavelet transformation and hashtags that present a peak in their pattern of occurrence are selected. After filtering the hashtag, LDA is used to infer topics associated with each tag. For intervals of five minutes, all the tweets containing the hashtag are retrieved and passed to the LDA algorithm for the computation of five representative topics of the hashtag. LDA has been modified in~\citet{Pan2011} in order to take into account temporal and geographic features. The approach in \citet{Pan2011} represents a pioneering work, but it suffers from the fact that either the number of topics or the number of events has to be fixed. 

\citet{Angel2014} develop an algorithm for maintaining overlapping dense subgraph with size limit in a dynamic graph. Each dense subgraph represents an event, where the nodes are words describing the event. Their main contribution consists in designing an efficient data structure for maintaining overlapping subgraphs.

\citet{Guille2014} leverage the intuition that during an event users will mention more often other users in order to engage in conversation or they will retweet posts of other users. The words from tweets containing mentions are selected and for each of them, the deviation from the expected frequency for given time periods is computed. To provide a better description of an event, additional words sharing similar temporal patterns with the initial word are added.

Event detection shares some similarities with trend detection,\footnote{\url{https://blog.twitter.com/2015/trend-detection-social-data}} which is one main focus for Twitter. Trend detection aims at finding keywords or hashtags whose frequency deviates significantly from the average frequency. However, it does not focus on a given class of real-world events (such as disaster events), nor does it provide a short description of the topics.

We evaluate our algorithm against~\citet{Guille2014}, as it is the approach which performed best in our experimental evaluation. We select also the technique presented in~\cite{WengL11}, as it has been shown~\cite{Weiler2015} to perform better than several other event detection techniques~\cite{Weiler2013,Weiler2014,Cordeiro2012}. 

\textbf{Dense subgraphs.} Dense subgraphs discovery has applications in many areas, for example in community detection~\cite{Fortunato2010}, finding patterns in gene annotation graphs~\cite{Saha2010}, link spam detection~\cite{Gibson2005} and event detection~\cite{Angel2014}. One of the most well known and studied definitions of density is the average degree density and the problem of finding a subgraph that has the maximum average degree is called the densest subgraph problem. Goldberg \cite{Goldberg1984} formally defined the problem in an undirected graph and presented an algorithm that computes the densest subgraph in $O(log(n))$ maximum-flow computations.  
When enforcing a limit on the size of the subgraph, finding the densest subgraph of exactly k vertices $(DkS)$, the problem becomes NP-hard. Variations on the problem, finding a densest subgraph with at most k vertices$(DamkS)$ or a densest subgraph with at least k vertices $(DalkS)$ are also NP-hard \cite{Khuller2009}. Recent work~\cite{letsios2016}, showed that exact solutions for the heaviest $k$-subgraph ($DkS$ in a weighted graph) can be obtained efficiently when considering real-world graphs.  
An interesting variation on finding a dense subgraph is the cocktail party problem~\cite{sozio10}, where the task is to find a dense subgraph that contains a specified set of input nodes.

Cliques and quasi-cliques are dense subgraphs par excellence. Given an undirected, unweighted graph, a clique is a fully connected subgraph. A quasi-clique in an unweighted graph has been defined as a subgraph with a number-edges-to-number-edges-clique-same-size ratio larger than a given threshold value in~\cite{Uno2010}, and a subgraph maximizing the edge surplus
over the expected number of edges under the random-graph model in~\cite{BabisKDD13}. When considering undirected weighted graphs, cliques have been defined either as a subgraph of maximum total weight where any two nodes are connected~\cite{Ostergard}, or as a subgraph with a sum-edge-weight-to-number-edges-clique-same-size ratio larger than a threshold~\cite{Uno2010}.

\section{Algorithms}\label{sec:algo}

%

Our algorithm consists of the following main steps: 1) collection of tweets containing keywords related to disaster events by means of the Twitter API; 2) recognition and tagging of mentions of locations in the tweets; 3) finding bursts of mentions of locations; 4) mentions of locations are finally complimented with related keywords so as to provide more informative results. Each of these steps is described in the following sections.


\subsection{Collection of Tweets and Preprocessing}\label{sec:eventspreproc}

Tweets have been collected by means of the Twitter API while specifying a list of keywords related to disaster events, such as \textit{attack}, \textit{flood}, \textit{victims}. To this end, we use the list of keywords provided in~\cite{Olteanu2014}. For the recognition and tagging of locations, we use an entity recognition tagger that was trained on Twitter data \cite{Ritter2011}. Such a tagger focuses on ten different categories: person, location, company, product, facility (e.g. Cornell University), tv show, movie, sports team, and band. We retain only location and facility tags while ignoring the others. Our intuition is that bursts in the mentions of a location ( in tweets dealing with disaster events ) might signal the happening of an important event in that location. After the tagging step, we lemmatize the words and then we filter them such that to remove stop words, URLs and infrequent terms (i.e. terms with an hourly frequency smaller than 5).

\subsection{Finding Bursts of Locations}\label{sec:eventsburst}

When an event such as a disaster event occurs, we observe a burst of activity in Twitter with terms pertinent to the event increasing suddenly their frequency in tweets. In our approach (where tweets contain keywords related to disaster events), a burst in the number of mentions of a location gives us a first signal that a disaster event is unfolding at that location. Previous works~\cite{Pan2011} have used geolocation of posts and not mentions of locations. We argue that using mentions of locations makes our algorithm more versatile, allowing it to analyze datasets coming from different sources.
Burstiness of words in streams of data is a well-studied topic~\cite{zhu2003,Lappas2009} and in our approach, we use a simple technique, similar to other event detection methods~\cite{Guille2014}. 

For each location, we compute a set of intervals in which the deviation between the frequency of the location and its expected frequency is always above a threshold. Our intuition is that all tweets (dealing with the same location) posted during each of those intervals refer to the same event. We refer to such intervals as \emph{interesting intervals}. We are interested in finding \emph{maximal} interesting intervals. The expected frequency of a location is computed assuming that location frequencies can be approximated by the binomial distribution.
Our algorithm computes for every location and every maximal interesting interval of that location its frequency and how much it deviates from the expected frequency. Then, all the (location, maximal interesting interval) pairs are ranked according to how much the frequency of a location deviates from its average frequency. A larger deviation corresponds to a higher interest in the event, therefore we retain the top $k$ (location, maximal interesting interval) pairs with the largest deviation from the average. 

Below we provide more technical details. The expected frequency of a term (a location in our case) is computed as follows. Let $X_w$ be the random variable indicating the number of tweets containing a term $w$ while letting $N$ be the number of tweets. We assume that the probability of having $n$ tweets containing $w$, denoted as $P(X_w=n)$, can be approximated by a binomial distribution:

\[ P(X_w = n)  = \binom{N}{n} p_w^{n} (1-p_w)^{N - n},\] 

where $p_w$ is the probability of a tweet containing the term $w$. In order to have an estimate of $p_w$, we sample a collection of tweets written well before the event (we will take tweets from the previous month). Given the percentage of tweets containing $w$ in that collection, we obtain a $95\%$ interval of confidence for $p_w$ using Wilson score method~\cite{newcombe1998two}. 


Afterwards, we can compute the expected number of tweets containing a term $w$ and the standard deviation as follows:

\[ E[X_w]  = N p_w \quad \textrm{and}  \quad \sigma[X_w] = \sqrt{ N p_w (1 - p_w)}.\] 

Let $f_{l,w}(t)$ be the frequency of a term $w$ in all tweets posted in the time window $[t,t+l]$. We let \[ \beta_{l,w}(t) = \frac{f_{l,w}(t) - E[X_w]}{\sigma[X_w]}, \] which measures how much $f_{l,w}(t)$ deviates from the expected frequency of the term $w$. We say that an interval $\mathcal{I} = [a,b]$ is \emph{interesting} with respect to a term $w$ if for all $t \in \mathcal{I}$, $\beta_{l,w}(t) \geq \alpha$. We say that an interesting interval $\mathcal{I}=[a,b]$ is maximal if for all $\bar{a}<a$ and $\bar{b}>b$, $[a,\bar{b}]$, and $[\bar{a},b]$ are not interesting.

Given a maximal interesting interval $\mathcal{I}$, we define $\beta_{l,w}^{\max}(\mathcal{I})=\max_{t \in \mathcal{I}}\beta_{l,w}(t)$, i.e. the maximum deviation from the expected frequency of $w$ in $\mathcal{I}$. When it is clear from the context we use the abbreviation $\beta_{\max}$. 
Our goal is to find all maximal interesting intervals for every term $w$ together with their corresponding $\beta_{max}$'s. We then retain only the top $k$ intervals with maximum $\beta_{max}$'s among all the terms. 
Algorithm~\ref{algo:burst} shows a pseudocode for computing all bursts of a given location, while Algorithm~\ref{algo:events} shows a pseudocode for computing the top $k$ maximal interesting intervals together with their corresponding locations, as well as other relevant terms (to be discussed next).



  
\begin{algorithm}[H]
\caption{FindTermBurst$(\mathcal{C}, w, \alpha, \ell)$}
\label{algo:burst}
\begin{algorithmic}[1]
\State \textbf{Input}: A collection of tweets $\mathcal{C}$ with their timestamps, a location $w$, a threshold $\alpha \geq 0$, a window size $\ell$
\State $S \gets \emptyset, \beta_{max} \gets 0, \mathcal{I} \gets \emptyset$
\State $t \gets minTimestamp(\mathcal{C})$
\While{$ t \leq maxTimestamp(\mathcal{C}) - \ell$}
\If{$\beta_{\ell,w}(t) \geq \alpha$}
\State $\mathcal{I} \gets \mathcal{I} \cup [t,t+\ell]$
\State $\beta_{\max} \gets \max (\beta_{\max},\beta_{\ell,w}(t) )$
\EndIf
\If{($\beta_{\ell,w}(t)  <  \alpha$ or $t=maxTimestamp(\mathcal{C}) - \ell$) and $\mathcal{I} \neq \emptyset$}
\State $S \gets S \cup (w,\mathcal{I},\beta_{\max})$
\State $\mathcal{I} \gets \emptyset$, $\beta_{max} \gets 0$
\EndIf
\State $t \gets t+1 $
\EndWhile
\State Return $S$
\end{algorithmic}
\end{algorithm}

Locations in tweets might be mentioned at different levels of granularity (city, state or a country). As a result, two tweets might refer to the same event, even if two different locations are mentioned in the corresponding tweets (e.g. Los Angeles and California). Therefore, determining whether there is a hierarchical relationship between two locations might help us determining whether two events are duplicate events, thereby improving the fraction of unique events. To this end, we use GeoNames, a geographical dataset which can be queried using web services~\footnote{\url{http://www.geonames.org/export/web-services.html}}.  In our algorithm we remove a tuple $(l^{'},\mathcal{I^{'}},\beta_{\max}^{'})$ from further consideration, if there exists another tuple $(l^{''},\mathcal{I^{''}},\beta_{\max}^{''})$ with $\mathcal{I^{'}} \cap \mathcal{I^{''}} \neq \emptyset, \beta_{\max}^{'} \leq \beta_{\max}^{''}$ and an hierarchical relation between $l^{'}$ and $l^{''}$.

\subsection{Finding Quasi-Cliques}\label{sec:eventsaggregation}
In order to complement the set of locations (found with Algorithm~\ref{algo:burst}) with additional information about the corresponding event, we employ a graph mining approach. In particular, for each location and each interesting interval for that location, we wish to find a set of terms which induce a dense region in the co-occurrence graph during that time interval. Given an interesting interval $\mathcal{I}$ and a collection of tweets, we define a weighted undirected graph $G_{\mathcal{I}}=(V_\mathcal{I},E_{\mathcal{I}})$, where $V_\mathcal{I}$ consists of the set of terms in the collection of tweets, while there is an edge between two nodes if the corresponding terms co-occur in at least one tweet posted within $\mathcal{I}$. A weight function $c:E \rightarrow \mathbb{R}^+$ represents the number of co-occurrences of terms in tweets posted within $\mathcal{I}$.

\begin{figure}[H]
\centering
\includegraphics[width=0.5\textwidth]{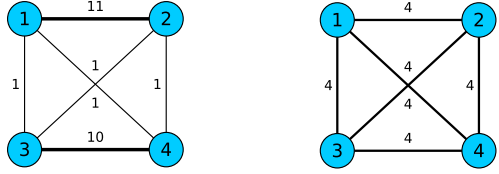}
\caption{The left subgraph has a larger weight than the right one, but the nodes in the right subgraph are much better connected.}
\label{fig:cliques}
\end{figure}

Cliques and quasi-cliques are dense subgraphs par excellence. Several definitions of weighted cliques have been provided in the literature, such as a subgraph of maximum total weight where any two nodes are connected~\cite{Ostergard}, as well as a subgraph with a sum-edge-weight-to-number-edges-clique-same-size ratio larger than a threshold~\cite{Uno2010}. Observe that both definitions would favor the graph on the left in Figure~\ref{fig:cliques}, which exhibit weak connections between the set of nodes $\{1,2\}$ and \{3,4\}. As a result, those two different parts of the graph might actually refer to two different events. It is more likely that the nodes of the graph on the right in Figure~\ref{fig:cliques} refer to the same event, as the edges of the graph have the same weight.

This motivates the following definitions of cliques and quasi-cliques. We say that a graph $H$ is a \emph{weighted clique} if all pairs of nodes in $H$ are connected by an edge with the same weight. Given a parameter $\gamma > 0$, we then define a \emph{weighted quasi-clique} as follows.

\begin{definition}[Weighted Quasi-Clique] Given an undirected weighted graph $H=(V(H),E(H), w)$, $0<\gamma \leq 1$, we say that $H$ is a \emph{weighted} $\gamma$-\emph{quasi-clique} if the following holds: \begin{equation}
 \sum_{e \in E(H)} w(e) \geq \gamma \cdot w_{max}(H) \binom{|V(H)|}{2}, 
\end{equation}
where $w_{max}(H) = \max_{e \in E(H)}(w(e)) $.
\end{definition}

We define the function $q_G:V \rightarrow (0, 1]$ ($q$ for short) to be the function which associates to every set $S \subseteq V$ a rational number $\gamma$ such that the subgraph induced by $S$ in $G$ is a $\gamma$-quasi-clique and $\gamma$ is the largest value for which this holds. 

Finding quasi-cliques is an NP-hard problem, therefore we resort to the following heuristic for finding quasi-cliques containing a node $v$ and at most $s$ nodes. The algorithm starts with $v$ and adds the edge with maximum weight containing $v$. At any given step, let $S$ be set of current nodes. If $|S|=s$, the algorithm stops. Otherwise, it adds a node $x$ in the neighborhood of $S$ maximizing $q(S \cup \{x\})$ provided that by adding $x$ the resulting subgraph is still a $\gamma$-quasi-clique. Algorithm~\ref{algo:quasi} shows a pseudocode, where $\Delta(u)$ denotes the sets of neighbors of $u$. 


\begin{algorithm}[H]
\caption{FindQuasiClique($G,v,\gamma$, s)}
\label{algo:quasi}
\begin{algorithmic}[1]
\State \textbf{Input}: A graph $G=(V,E)$, a vertex $v$, $0<\gamma \leq 1$
\State \textbf{Output}: A weighted quasi-clique containing $v$
\State $u \leftarrow \arg \max_{u: uv \in E} w(uv)$
\State $S \leftarrow \{u,v\}$
\While{true}
\State if $|S|=s$ return $S$
\State $\Delta(S) \gets \cup_{u \in S} \Delta(u) \setminus S$
\If{$\Delta(S) = \emptyset$}
\State return $S$
\EndIf
\State $x \leftarrow \arg \max_{u \in  \Delta(S)}q(S \cup \{u\})$
\If{$q(S\cup \{ x \}) < \gamma$}
\State return $S$
\EndIf
\State $S \leftarrow S \cup \{ x \}$
\EndWhile
\State Return $S$
\end{algorithmic}
\end{algorithm}


\subsection{\textsc{EviDense}: Analysis of its Running Time}\label{sec:eventsmerge}

Our algorithm for event detection can then be recapped as follows. The first step consists of collecting tweets from Twitter specifying as a filter a list of terms related to disaster events. Then, mentions of locations in tweets are recognized and tagged.  A list of the top $k$ bursty locations is computed, each of them is then complemented with additional terms related to the same event by finding quasi-cliques in the co-occurrence graph. Algorithm~\ref{algo:events} shows a pseudocode for our main algorithm.


\begin{algorithm}[H]
\caption{\textsc{EviDense}($\mathcal{C},\gamma, \alpha, \ell, k, s$)}
\label{algo:events}
\begin{algorithmic}[1]
\State \textbf{Input}: A collection  $\mathcal{C}$ of tweets each one containing at least one term related to disaster events (lexicon provided in~\cite{Olteanu2014}),
$0< \gamma \leq 1$, an integer $\alpha$, a window size $\ell$, integer $k,s$
\State \textbf{Output}: The list of the top $k$ events in $\mathcal{C}$
\State $S \gets \emptyset$
\State Recognize and tag locations in $\mathcal{C}$ with~\cite{Ritter2011}
\For{ $w \in Locations(\mathcal{C})$ }
\State $S \gets S \cup$ FindTermBurst$(\mathcal{C}, w, \alpha, \ell,k)$
\EndFor
\State Let $S_k$ be the top $k$ triples in $S$ with maximum $\beta_{\max}$'s
\For{ $(w,\mathcal{I},\beta) \in S_k$  }
\State Compute $G_{\mathcal{I}}$
\State \textbf{output} (FindQuasiClique ( $G_{\mathcal{I}},w, \gamma, s ), \mathcal{I}$)
\EndFor
\end{algorithmic}
\end{algorithm}



To ease the presentation, Algorithm~\ref{algo:burst} and Algorithm~\ref{algo:events} show a simple pseudocode of the main steps of our algorithm. However, our algorithm can be efficiently implemented so that it requires a few constant passes over the collection of tweets. For each location we store its frequency in the current time window (i.e. $f_{l,w}(t)$) together with the corresponding $\mathcal{I}, \beta_{\max}$. Every time we process a new tweet, we verify for every location in that tweet the conditions from Algorithm~\ref{algo:burst} lines 5-10. Let $c$ be the maximum number of terms in a tweet. We can then claim that the total running time of Algorithm~\ref{algo:burst} is $O(c|\mathcal{C}|)$.


Let $n,m$ be the number of nodes and the number of edges in $G$, respectively (i.e. the input of Algorithm~\ref{algo:quasi}). For line 3 we require $O(n)$ operations, for line 7 we need at each iteration $O(n)$ operations, while the computation at line 10 requires in the worst case to process all the edges in $G$. Thus, the overall running time of Algorithm~\ref{algo:quasi} is $O(s(m+n))$. We now evaluate the running time of Algorithm~\ref{algo:events}, lines 5-10, that is, excluding the tagger. Line 7 is implemented using a heap of size $k$, requiring a number of operation of $O(|\mathcal{C}|c \log k)$. Any graph constructed from a collection of at most $|\mathcal{C}|$ tweets will have at most $c^2 |\mathcal{C}|$ edges and $c |\mathcal{C}|$ nodes. For computing a graph $G_{\mathcal{I}}$, we need at most $O(c^2 |\mathcal{C}|)$ operations. Then, the running time for lines 8-10 is $O(ksc^2 |\mathcal{C}|)$, which implies a total running time of Algorithm~\ref{algo:events} of $O(T+ksc^2 |\mathcal{C}|)$, where $T$ is the running time required by the tagger. Notice that in practice $k,s,c$ are small constants. 


\subsection{Incorporating Information from Mainstream Media}
\label{sec:eventsnews}

By running \evidense\ on a collection of titles of news articles, we can detect high-impact events in mainstream media as well. This observation leads to the question of whether our approach could benefit from incorporating news articles from mainstream media. 

This is done as follows. Recall that the ranking of a result is determined by the deviation from the expected frequency of its location in a given interval. We introduce a new random variable $Y_{loc} = X_{loc}^{news} + X_{loc}^{tw}$, where $X_{loc}^{news}$ and $X_{loc}^{tw}$ are  random variables denoting the frequency of the location $loc$ in news and tweets in a given interval, respectively. We make the simplifying assumption that the random variables $X_{loc}^{news}, X_{loc}^{tw}$ are independent. Such an assumption holds up to some extent, given that news articles and tweets are written typically by different people. Therefore, the expected value and the standard deviation of $Y_{loc}$ can be computed as the sum of the expected values and standard deviations of the two random variables, respectively. We will then compute the maximum deviation of that location considering the sum of mentions of the location in both datasets.

To summarize, we first select the top $k$ bursty locations in mainstream media. For each of these locations, we shall consider the random variable $Y_{loc}$ to determine the expected frequency, while for the remaining locations we shall consider the variable $X_{loc}^{tw}$. We then compute the top $k$ results in Twitter.

\section{Experimental Evaluation}\label{sec:eval}

\subsection{Experimental Settings}\label{sec:exp}

We evaluate our approach on a large collection of tweets, as well as, news articles, by means of a crowdsourcing platform. We mostly focus on disaster events, while in Section~\ref{sec:beyond} we investigate if our approach is effective for other classes of events, as well. We consider a disaster, any event that is included in any of the lists provided by the US Government\footnote{\url{https://www.ready.gov/be-informed}} or the International Disaster Database\footnote{\url{http://www.emdat.be/classification}} (such as earthquakes, mass shootings, floods, etc.).

\noindent\textbf{Corpora.} We collect tweets posted over a period of 19 months between November 2015 and May 2017. We use the Twitter Streaming API while filtering the tweets so that they contain at least one term related to disasters~\cite{Olteanu2014}. We focus on tweets written in English. We obtain 16M tweets in total, which we divide into nineteen datasets (one per month). We then use an entity recognition tagger \cite{Ritter2011} for recognizing and tagging the mentions of locations in the tweets. We collected news articles over the same time period using GDELT\footnote{\url{http://blog.gdeltproject.org/gdelt-2-0-our-global-world-in-realtime/}}, as in~\cite{olteanu2015}, which contains major international, national, regional, and local news. The GDELT dataset is updated regularly, every 15 minutes, with news posted in that timeframe. We filtered the news using the same lexicon used to filter the tweets, which can be done by specifying the tag ``CRISISLEX\_CRISISLEXLEC''\footnote{\url{https://blog.gdeltproject.org/crisislex-taxonomies-now-available-in-gkg/}} in GDELT. From each news article, we retain its title and publication date which are given in input to our algorithm. In order to get an estimate of the probability of a tweet or a news article to contain a location, we sampled the month previous to the event, and we obtained a 95\% interval of confidence using Wilson score method\cite{newcombe1998two}. For simplicity, we selected the upper bound of the interval as the value of the probability.

\noindent \textbf{Related work.} We compare against \mabed~\cite{Guille2014} and \edcow~\cite{WengL11}. All approaches are evaluated on the same collection of tweets. There has been significant disagreement among the crowd workers when interpreting the results of~\cite{Angel2014}. Therefore, we omit the approach from our study, deferring a more careful evaluation to future work. 

 
\noindent \textbf{Parameter settings.} In our approach, we set $s$ to 10, $k$ to 20, $\ell$ to three hours, $\alpha = 8$ and $\gamma = 0.5$. 
We run \mabed\ using the implementation provided by the authors\footnote{\url{https://github.com/AdrienGuille/MABED}} and the setting specified in the original paper \cite{Guille2014}, that is $p = 10, \theta = 0.7$ and $\sigma = 0.5$. The parameter $\sigma$ controls the similarity between the events produced in the results. Small values of sigma correspond to more diverse results. We observe that determining the right value for such a parameter is crucial in the \mabed\ approach. In particular, if $\sigma$ is too small one may not get enough relevant results, while with large values of $\sigma$ the algorithm might return many duplicates of the same ``popular'' event. For the \edcow\ algorithm we use the implementation of \cite{Weiler2015} and we set the parameters as follows: the size of first level of intervals is $s = 100s$, while we take $\Delta = 32$, setting a size of $3200s$ for the second-level intervals and, as in \cite{Weiler2015}, we set $\gamma = 1$. As \edcow\ does not enforce any constraint on the size of the output, we order the results according to $\epsilon$ (as defined in the original paper), which measures the relevance of the results and retain only the top $k$ results.

\noindent \textbf{Machine stats.} We run our experiments on a Linux machine equipped with 4 processors Intel Xeon CPU E7-4870 @ 2.40 GHz as well as 10 cores split into 2 threads each (a total of 80 threads) and 64 G of RAM DDR3 1333 MHz. 

\subsection{Methodology}\label{sec:meth}

We run the three approaches on each of the first four months of our dataset from Twitter and evaluate the top 10 as well as the top 20 results for each such dataset. \mabed\ and \edcow\ provide a measure of relevance for their results which we use to determine the top results. The approach that performs best is then evaluated against our approach, over the whole period of 19 months. All approaches receive in input the same collection of tweets.

\noindent\textbf{Metrics.} We evaluate the precision at $k$ for all three approaches on our collection of tweets. Precision at $k$, denoted with $P@k$ (or precision for short), is defined as the percentage of true events in the top $k$ results, divided by $k$. In addition to the precision, we compute the percentage of duplicate events among all the events retrieved, i.e. the \textit{DeRate}~\cite{li2012}. From these two metrics, we can infer a third one, which measures the percentage of unique events (i.e. duplicates do not contribute) in the top $k$ results. We denote such a metric with $U@k$. We observe that if an algorithm performs best in terms of $U@k$, it performs best also in terms of recall. Therefore, we do not report recall in our experimental evaluation.

\noindent\textbf{Crowdsourcing Settings.} In order to ensure a fair comparison, we use a crowdsourcing service,  GetHybrid\footnote{\url{https://www.gethybrid.io}}. For each result produced by any of the approaches, we ask 5 workers to determine whether it was a disaster event, that is an event considered to be a disaster by the US Government or the International Disaster Database.\footnote{\url{https://www.ready.gov/be-informed}, \url{http://www.emdat.be/classification}}.  We added to the description of a result two relevant tweets in order to facilitate the labeling task. We note that as we find the relevant tweets automatically, a larger collection of tweets could be used for other applications, such as to create training sets for automatic classifiers. 

In order to evaluate a result, a worker would select one of the following answers to the question of what type of event do the keywords and tweets describe: 
\begin{description}

\item{(A)} A natural disaster (earthquake, landslide, volcano, extreme temperature, hurricane, large and dense fog, large storm, flood, tsunami, drought, wildfire, epidemic, insect infestation, large accident involving animals, asteroid impact), 

\item{(B)} A technological disaster (chemical spill, building collapse, explosion, fire, gas leak, large poisoning, nuclear, radiation, cyberattack),

\item{(C)} A human-induced disaster (war, shooting, terror attack), 

\item{(D)} A large transport accident (air, road, rail, water),

\item{(E)} Not a natural, technological, or human-induced disaster, or large transport accident.
\end{description}

Answers A-D correspond to a disaster.  We label a result as a disaster if 3 out of 5 workers confirmed. 

In order to compute the \textit{DeRate}, we use the following definition for duplicate events: two events (E1, E2) are duplicates if they can be referenced to a more general event (E), that is each event satisfies one of the conditions: 
\begin{enumerate}

\item describes the event E,

\item describes a sub-event of the event E, e.g. the Bataclan attack was one of the sub-events of the Paris attacks in November 2015,

\item describes the consequences of the event E, e.g. actions taken in the aftermath of a tornado.
\end{enumerate}

As before we label two events as being duplicates if 3 out of 5 workers confirmed. 

\noindent\textbf{Estimating \textit{DeRate} and U@k}. The task of estimating the average number of duplicate events is non-trivial, given the large number of results. Asking the workers in GetHybrid to estimate the number of duplicate events in a list containing approximately $20$ or more results is time-consuming and most probably would result in a non-accurate evaluation.  Therefore, we draw a random sample from the set of all possible event pairs. Each worker is then asked to determine whether a given pair of events in the sample contains duplicate events or not. The resulting percentage of duplicate event pairs in the sample is used to infer a $95\%$ confidence interval on the percentage of duplicate event pairs on the whole dataset, using the Wilson score method~\cite{newcombe1998two}.

We then proceed to estimate the number of unique events produced in output by the different approaches. Given a $95\%$ confidence interval  $[a,b]$ for the number of duplicate event pairs, we compute a $95\%$ confidence interval $[\bar{a},\bar{b}]$ for the number of duplicate events as follows. In the case when there are at least $a$ duplicate event pairs, there are at least $\bar{a}=n-1$ duplicate events where $n$ is the smallest integer such that $\binom{n}{2} \geq a$. If there are at most $b$ duplicate event pairs, there can be at most $\bar{b}=b$ duplicate events. Therefore, for an approach with precision (at $k$) $p>0$ and $95\%$ confidence interval $[\bar{a},\bar{b}]$ on the number of duplicate events, we estimate that the number of unique events be in the range $[\max (1,p\cdot k- \bar{b}) , p\cdot k-\bar{a}]$.

\subsection{Results}
\label{sec:results}
\noindent\textbf{Comparison.} In Table~\ref{tab:avgprec} we present the average $P@10$ and the average $P@20$ for all the approaches over the time period November 2015 - February 2016. The evaluation has been conducted using the GetHybrid crowdsourcing platform, as discussed in Section~\ref{sec:meth}. We observe that \evidense\ outperforms the other two approaches for both top-10 and top-20 results. 

\begin{table}[!htbp]
\small
\centering
\begin{tabular}{|c|c|c|}
\hline
Method & Average $P@10$ &  Average $P@20$  \\
\hline
\edcow & 42.5\% &    32.5\%\\
\mabed  & 60\% &    53.7\%\\
\evidense & 70\% &    73.7\%\\

\hline
\end{tabular}
\\~\\
\caption{Average precision over the time period November 2015 - February 2016.}
\label{tab:avgprec}
\end{table}

We evaluate the number of duplicate events in the results, as discussed in Section~\ref{sec:meth}. First, we measure the percentage of duplicate event pairs in our sample. This is shown in Table~\ref{tab:95CI}. Observe, that \evidense\ produces less duplicate event pairs. In particular, our results are better with a $95\%$ confidence.

\begin{table}[!htbp]
\centering
\small
\begin{tabular}{|c|c|c|}
\hline
Method & Percentage in sample & $95\%$ Confidence Interval \\
\hline
\edcow & 12\% & 4.9\% to 25\%\\
\mabed  & 22\% & 14.5\% to 31.6\%\\
\evidense & 2\% & 0.3\% to 7.7\%\\
\hline
\end{tabular}
\\~\\
\caption{Percentage of duplicate event pairs.}
\label{tab:95CI}
\end{table}

From the results shown in Table~\ref{tab:95CI}, we obtain a $95\%$ confidence interval on the percentage of duplicate events, i.e. the \textit{DERate}. We obtain a $95\%$ confidence interval of $[0\%, 84.6\%]$ for \edcow,  $[37.2\%,90.6\%]$ for \mabed, and  $[0\%, 47.4\%]$ for \evidense. From the latter result, it is difficult to determine which algorithm performs best in terms of \textit{DERate}. Moreover, observe that approaches with higher precision might be penalized by the \textit{DERate}, in that, they tend to have a larger number of duplicates. For example, an approach which retrieves exactly one event has a \textit{DERate} of zero. Therefore, we also consider the $U@k$ metric, that is, the percentage of unique events in the top $k$ results, as discussed in Section~\ref{sec:meth}. The results are shown in  Table~\ref{tab:nodupl}. We can see that even with a pessimistic estimate, \evidense\ outperforms the other approaches in terms of $U@k$, while the percentage of unique events in the top-$20$ results can be up to $73.7\%$.


%
%

The output of our algorithm is shown in Table~\ref{tab:ustop10}. We can see that the description of each of the events is succinct and informative. In particular, one can easily retrieve the location of the event (in bold), its time-frame and what happened.


%
%
%
%

\begin{table}[!htbp]
\centering
\small
\begin{tabular}{|c|c|}
\hline
Method & $U@20$  \\
\hline
\edcow & 5\% to 32.5\%\\
\mabed  & 5\% to 33.7\%\\
\evidense & 38.7\% to 73.7\%\\
\hline
\end{tabular}
\\~\\
\caption{Average percentage of unique events over the time period November 2015 - February 2016.}
\label{tab:nodupl}
\end{table}

\noindent\textbf{Long-Term Comparison.} In order to give further proof of the performance of our approach, we compute the precision with $95\%$ confidence of \evidense\ and the second best algorithm \mabed\ for the period of 19 months of our dataset. We retrieve the top 20 results for every month, giving a total of 380 results for each approach. From these results, we randomly selected 100 results per method and we used crowdsourcing to evaluate the precision. For \evidense, 72 events out of 100 are disaster events, giving a precision between $61.9\%$ and $80.3\%$ with $95\%$ confidence, while for \mabed, 33 events out 100 are disaster events, resulting in a precision between $24.1\%$ to $43.2\%$ with $95\%$ confidence. These results are summarized in Table~\ref{tab:long}. This confirms that our approach has higher precision than \mabed, while it performs remarkably well over such an extensive experimental evaluation. 

\begin{table}[!htbp]
\centering
\small
\begin{tabular}{|c|c|c|}
\hline
Method & Percentage in sample & $95\%$ Confidence Interval \\
\hline
\mabed  & 33\% & 24.1\% to 43.2\%\\
\evidense & 72\% & 61.9\% to 80.3\% \\
\hline
\end{tabular}
\\~\\
\caption{Average precision over the period November 2015 - May 2017 for the best two methods.}
\label{tab:long}
\end{table}

\begin{table}[!htbp]
\small
\begin{tabular}{|c|c|c|}
\hline
Method&  Average $P@10$  &  Average $P@20$ \\
\hline
\evidense\ on News &    80\%    & 66.2\%    \\
\evidense\ on Twitter &    70\%    & 73.7\%    \\
\evidense\ Merged & 77.5\% &  76.2\%     \\
\hline
\end{tabular}
\\~\\
\caption{Comparison of the performance of \evidense\ on different datasets: news articles, Twitter and both.}
\label{tab:news}
\end{table}

\noindent\textbf{Tweets \& News Articles.} We investigate whether our approach could benefit from incorporating another source of information such as news articles from mainstream media, as discussed in Section~\ref{sec:eventsnews}. The results are shown in Table~\ref{tab:news}. We observe that we obtain significantly better results for top-10 and slightly better results for top-20. Overall, the precision improves when adding additional information from mainstream media.

\textbf{Running time.} Annotating the entities in the tweets using the entity recognition tagger \cite{Ritter2011} takes on average 0.02s/tweet. The average running time for the algorithms (without taking into account the annotating phase) is the following: for \mabed\  is around 90s, for \edcow\ is 52min and around 25s for \evidense.



  \begin{table*}[!htbp]

    \small
    \centering

    \begin{tabular}{|p{3.5cm}|p{5cm}|p{4cm}|}

    \hline

    Time interval (UTC) & Event keywords & Description \\ \hline

     \hline

    Dec 03 02:20 , Dec 07 08:50 & \textbf{San Bernardino}, dead, female, \#sanbernardino, killed, male, police, shooting&  The San Bernardino terrorist attack.\\ \hline

    Dec 27 01:04 , Dec 28 03:37 &\textbf{Dallas}, Rowlett, tornado  & Many tornadoes in Dallas and Rowlett. \\ \hline
    Dec 07 03:24 , Dec 10 21:11 &\textbf{Chennai}, damaged, flood, fresh, issue, lost, passport, psks, sushmaswaraj &  Chennai residents whose passports were damaged during the floods could get a new one.\\ \hline

    Dec 04 09:12 , Dec 04 23:23 &\textbf{Cairo}, attack, firebomb, killed, nightclub, people, restaurant & People killed in Cairo at a restaurant that operated also as a nightclub. \\ \hline

    Dec 28 20:15 , Dec 29 09:21 & \textbf{Cleveland}, 12-year-old, Tamir Rice charged, death, grand, indict, jury, police, prosecutor, shooting   & Not a disaster event. \\ \hline

    Dec 01 12:32 , Dec 06 22:03 &\textbf{Chennai}, find, flood, girl, lost, parent, pls & A girl has disappeared during the floods and her parents are looking for her. \\ \hline

    Dec 27 01:39 , Dec 27 08:38 &\textbf{Garland}, Texas, dead, hit, killed, people, storm & A series of tornados hit the state of Texas, reaching Garland, Dallas and Rowlett (duplicate event). \\ \hline
 
    Dec 21 04:56 , Dec 22 00:57 &\textbf{Las Vegas}, crash, critical, dozen, driver, injured, pedestrian, people, strip & Driver in Las Vegas deliberately attacked pedestrians. \\ \hline

    Dec 18 16:40 , Dec 19 02:32 & \textbf{republican}, agrabah, aladdin, bombing, country, \#nottheoni, nationally, ppppolls, primary& Not a disaster event, also we can notice a mislabelling of the tagger.\\ \hline

    Dec 07 13:31 , Dec 08 08:59 &\textbf{Pearl Harbor}, attack, \#pearlharbor, honor, live, lost, remember, today, year & Commemoration of the attack on Pearl Harbor 74 years ago today.\\ \hline

     \hline

    \end{tabular}
    \\~\\~\\
    \caption{Top 10 events discovered in December 2015 by \evidense. The event is centered on the location given in bold.}

    \label{tab:ustop10}

    \end{table*}

\subsection{Parameter Settings for \evidense}\label{sec:eventsparam}

\begin{figure*}[!htbp]
\centering
\begin{subfigure}[t]{0.45\textwidth}
\includegraphics[width=\textwidth]{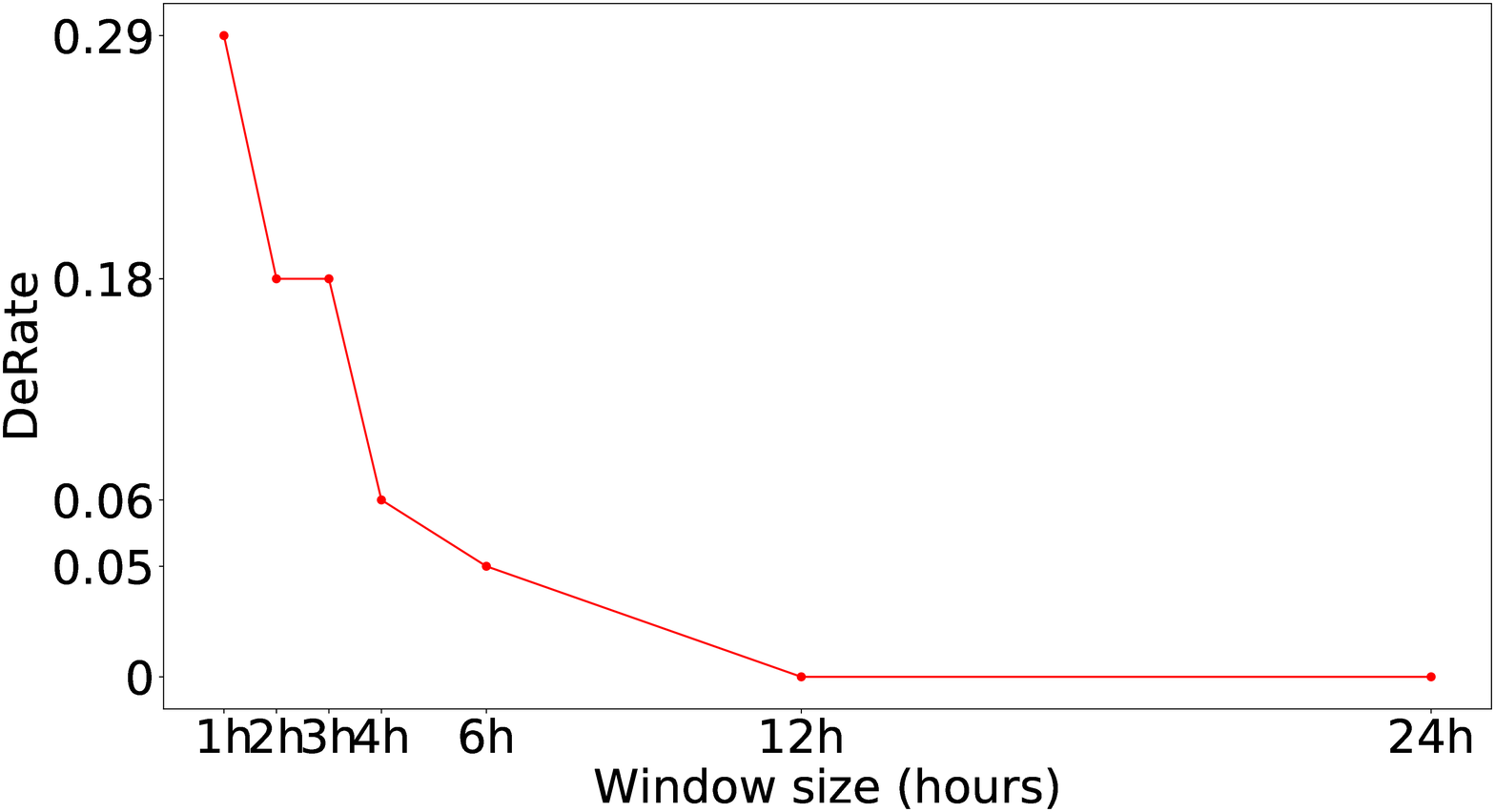}
\caption{DeRate vs $\ell$}
\label{fig:set1}
\end{subfigure}
\begin{subfigure}[t]{0.45\textwidth}
\includegraphics[width=\textwidth]{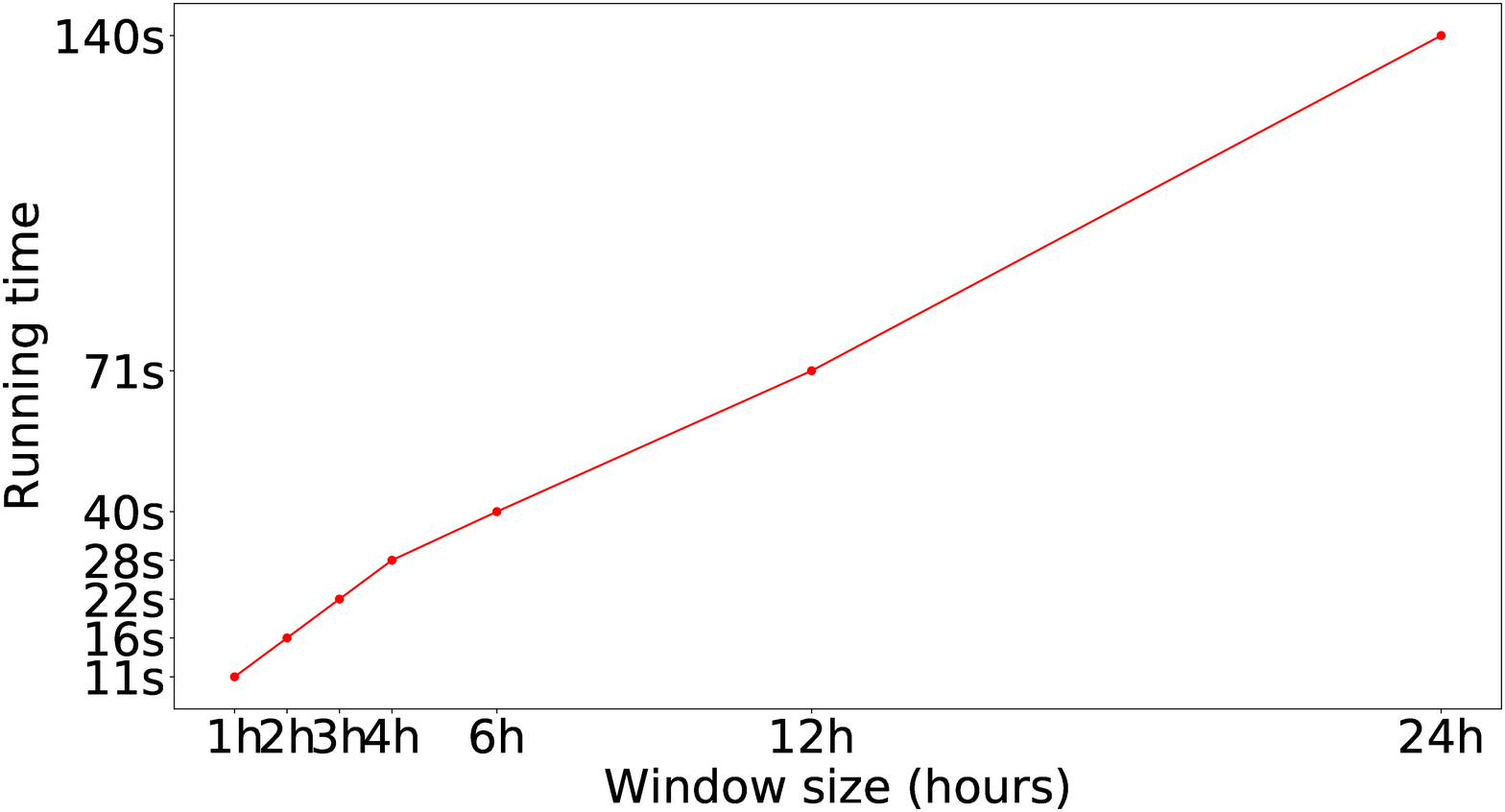}
\caption{Running time vs $\ell$}
\label{fig:set2}
\end{subfigure}
\begin{subfigure}[t]{0.45\textwidth}
\includegraphics[width=\textwidth]{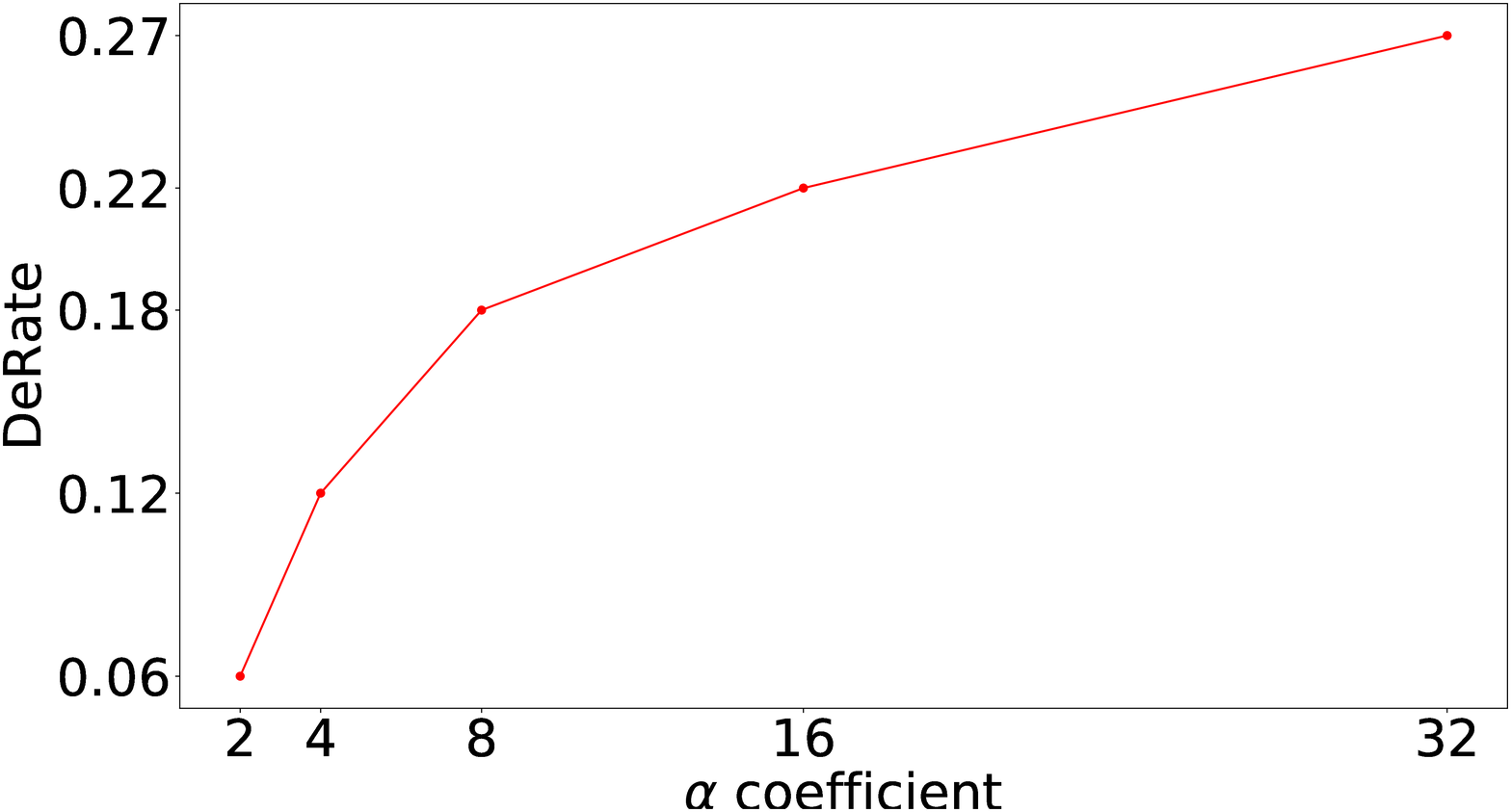}
\caption{DeRate vs $\alpha$}
\label{fig:set3}
\end{subfigure}
\begin{subfigure}[t]{0.45\textwidth}
\includegraphics[width=\textwidth]{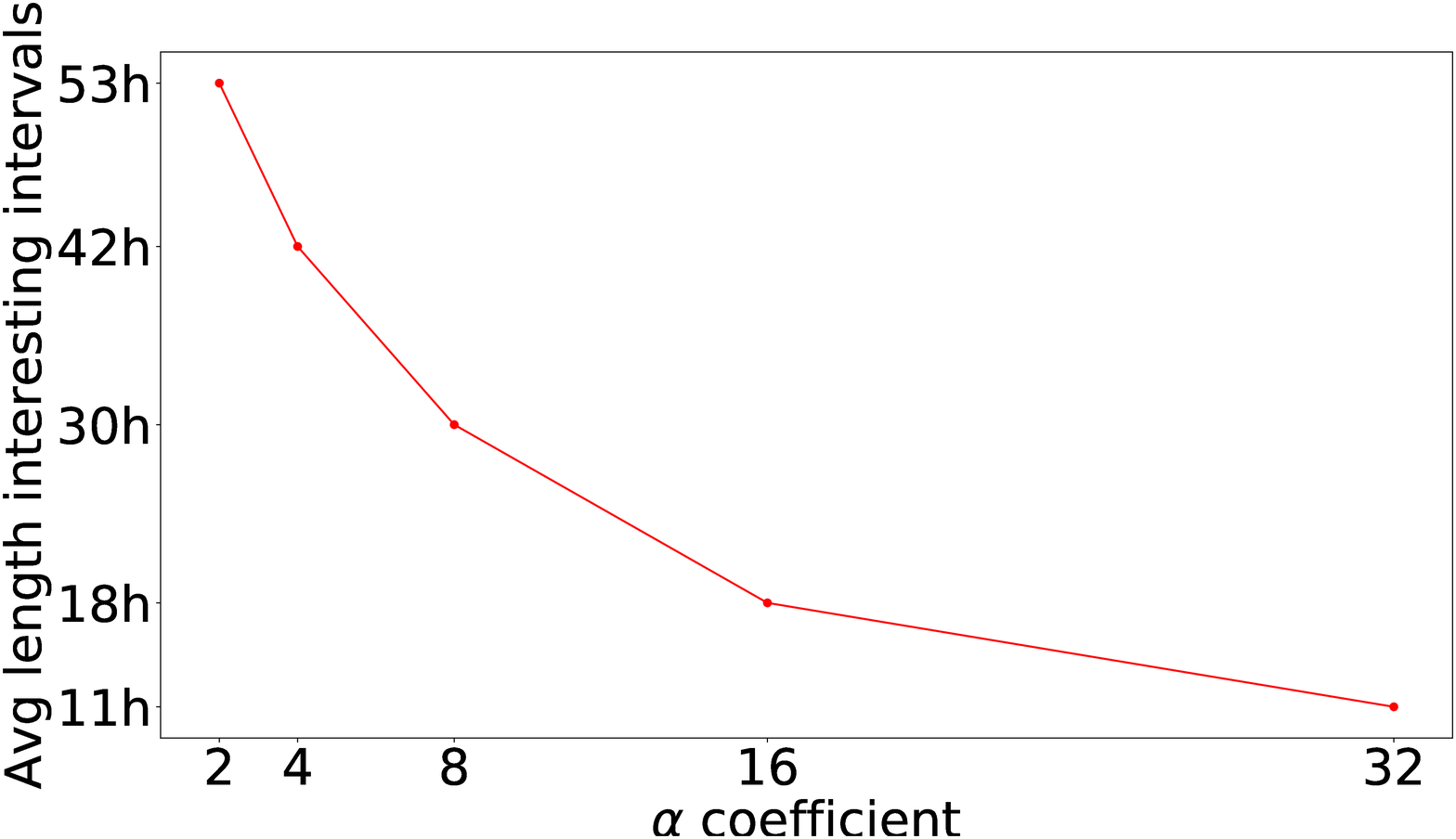}
\caption{Avg length of maximal intervals vs $\alpha$}
\label{fig:set4}
\end{subfigure}
\caption{Parameter settings}

\end{figure*}

Our algorithm has four input parameters: $k, \ell, \alpha$ and $\gamma$. For the experiments we used the values $k = 10$ or $k = 20$, we set $\ell$ to three hours, $\alpha = 8$ and $\gamma = 0.5$. The first parameter $k$ is used to limit the output size. The size of the time-window, $\ell$, has two functionalities: firstly, it has a smoothing effect over the number of occurrences of a term and can alleviate the difference between active periods and inactive periods, such as day and night. Secondly, because of the smoothing effect, it can decrease the importance of small events which have short-lived peaks and increase the importance of events which are discussed for longer periods but without having important peaks. We experiment with different values of $\ell$ in order to balance between the segmentation of a single event and the order of magnitude of an event. In order to assess what is the best value for the time window, we compute the percentage of duplicate events between the events retrieved, the \textit{DeRate}. We only include an evaluation of the top 20 events in the month of December 2015, as the results are similar to the other datasets. In Figure~\ref{fig:set1}, we can notice that the number of duplicate events decreases as the time window increases, which is intuitive, as a larger time window decreases the effect of an inactive period. We settled for three hours, as choosing larger time windows would likely merge different events that occur in the same location. Moreover, larger time windows would also increase the running time of the algorithm, as shown in Figure~\ref{fig:set2}. This is due to the fact that the search for weighted quasi-cliques is performed on a larger graph. 

The parameter $\alpha$ is used as a threshold for determining which location has a frequency which deviates significantly from the average frequency.
We use the binomial distribution as a simple model that gives an estimate of a mean and standard deviation, and we try several values for $\alpha$, as shown in Figure~\ref{fig:set3}. When increasing the parameter $\alpha$ the average length of interesting intervals decreases (Figure~\ref{fig:set4}) and given that our goal is to minimize the length of interesting intervals without reporting the same event twice, we select the value $\alpha=8$.
The last parameter, $\gamma$, is used to provide different levels of detail concerning an event. We have obtained good results for a large range of values of $\gamma$. The best is to avoid values smaller than $0.3$ which might add unrelated terms to an event, and values larger than $0.8$ as the description of the events could be not informative.

\subsection{Disaster Dataset}\label{sec:dataset}
We released a dataset~\footnote{\url{https://github.com/nyxpho/evidense}} containing 32K tweets written in the period November 2015 -February 2016. The dataset was obtained by selecting tweets referring to each true disaster event discovered by \evidense. The tweets were written in the timeframe of the disaster event and contained the mention of the event location and of at least one more event keyword. In order to test the quality of the dataset, we randomly sampled 200 tweets and we marked them as being disaster-related or not. A tweet was considered as being disaster-related if it described a disaster event, described a sub-event of a disaster event or described the consequences of a disaster event. We obtained a precision of $86.5\%$, which gives a $95\%$ confidence interval~\cite{newcombe1998two} for the precision in the entire dataset between $83.05\%$ and $92.42\%$.

\subsection{Beyond Disaster Events}\label{sec:beyond}

In this section, we study a different category of events in order to understand whether our approach can effectively be used to detect other classes of events. We focus on political events. In order to obtain tweets relevant to our task, we create a lexicon of terms related to politics. We include keywords like ``protest'', ``immigration'', ``election'' or ``government'', however, better results could be obtained if the list of words is produced in the same fashion as the lexicon provided in~\cite{Olteanu2014}.
Detecting bursts in the mentions of locations turned out to be an effective method to detect disaster events. In the case of political events, politicians might play an important role in this kind of events. Therefore, we detect bursts in the mentions of locations or persons, as the latter ones might refer to well-known politicians. In order to recognize and tag mentions of well-known persons, we use we use the same entity recognition tagger used in the rest of paper~\cite{Ritter2011}. The rest of the algorithm remains unchanged.

We obtain the following results. In the month of November 2015 the first event retrieved by our algorithm is related to the declaration of François Hollande that France will close its borders in the aftermath of the November 13th attacks described by the following terms ``attacks'', ``France'', ``borders'', ``closes'', ``declares'', followed by an event about people injured at protests calling for the resignation of South Korea's president (``Korea'', ``beaten'', ``happening'', ``people'', ``protest''). In December 2015 most of the detected events are related to the presidential elections in the US (``bernie'', ``hillary'', ``\#demdebate'', ``war''), however, we find also a mention of the killing of peaceful protesters during a pro-Biafra protest (nigerian, \#ipob, navy, police, protester, shooting, unarmed). In January 2016 the UN holds an emergency meeting after North Korea claims having done a nuclear test (North Korea, council, emergency, meeting, security, set, test), while in February 2016 Jeb Brush and Rand Paul suspend their presidential campaigns.  Another event detected in February is related to the Yemen civil war (Yemen, \#usakillyemenis, saudi-led, us-backed) and was triggered by a bombing at a cement factory close to the capital of Yemen, which resulted in the death of civilians. 

These observations suggest that our approach is robust and it has good potential in providing meaningful results for other classes of events. 

\section{Using Dense Subgraphs for Event Description}
\label{sec:dense}

In this section, we investigate if the weighted quasi-clique definition that we introduced in this paper is the best suited for describing events.  We compare our method with well-known definitions of dense subgraphs. 


\subsection{Dense subgraphs definitions and algorithms} 

In this section, we give the definitions for the types of dense subgraphs we consider for the comparison. We make simple modifications to some of the baselines, in order to enforce that a query node $q$ is part of the solution.

\begin{definition}[Heaviest $k$-subgraph containing query node $q$ (HkS)]
Given an undirected weighted graph $G = (V, E, w)$, an integer $k$ and a query node $q$, the induced subgraph $H = (V(H), E(H))$ is the heaviest $k$-subgraph containing query node $q$, if $H$ has size $k$, contains the query node $q$ and has maximum sum of edge weights. 
\end{definition}

\begin{definition}[Heaviest $k$-clique containing query node $q$ (HkC)]
Given an undirected weighted graph $G = (V, E, w)$, an integer $k$ and a query node $q$, the induced subgraph $H = (V(H), E(H))$ is the heaviest $k$-clique containing query node $q$, if $H$ is a $k$-clique, contains the query node $q$ and has maximum sum of edge weights.
\end{definition}

\begin{definition}[Cocktail party problem (CP)]
Given an undirected weighted graph $G = (V, E, w)$ and a query node $q$, the induced subgraph $H = (V(H), E(H))$ is a solution to the cocktail party problem if $H$ contains the query node $q$, is connected and maximizes the minimum weighted degree of all nodes in the subgraph. 
\end{definition}
This definition is a particular case of Problem 1 defined in~\cite{sozio10}.

\begin{definition}[Cocktail party problem with distance d (CPD)]
Given an undirected weighted graph $G = (V, E, w)$, an integer $d$ and a query node $q$, the induced subgraph $H = (V(H), E(H))$ is a solution to the cocktail party problem with distance $d$ if $H$ contains node $q$, is connected, all the nodes of $H$ are connected to the query node through a short path of at most size $d$, and maximizes the minimum weighted degree of all nodes in the subgraph. 
\end{definition}
This definition is a particular case of Problem 2 defined in~\cite{sozio10}.

Finally, we also consider the following simple baseline.
\begin{definition}[Top k co-occurrences (Topk)]
Given a query word $q$, a collection of tweets $C$ and an integer $k$, the top $k$ word co-occurrences represent $k$ words co-occurring the most often in a tweet with word $q$.   
\end{definition}




These four definitions of dense subgraphs are well studied in literature~\cite{aitt00, Ostergard, letsios2016, sozio10}. In order to compute the dense subgraphs, we implemented the techniques presented in \cite{letsios2016, sozio10} with small adjustements for our problem. We describe these implementations in the following.
In \cite{letsios2016}, the authors present an exact and an approximate algorithm for the heaviest $k$-subgraph. The algorithm is a well designed branch and bound technique that enumerates the candidate solutions. The approach is bottom up, that is it gradually adds nodes to the candidate solutions. We enforce that the query node is present in the final solution by adding it in the initialization of the candidate solutions. 

In order to compute the heaviest $k$-clique containing a query node $q$, we modify the branch and bound algorithm \cite{letsios2016}. We shortly remind the approach: in the first step the edges are sorted in non-increasing order of the weights and each edge has assigned its index in the order. We can visualize the branch and bound algorithm as a tree, in which the root corresponds to all possible solutions and each node to a subset of solutions. Each node on a level $i$ has two children, one child corresponding to the solutions containing edge $i$ ( where $i$ is the position in the sorted order) and one to the solutions that don't contain the edge. A node has an upper bound which represents the maximum possible weight of a solution of that node and a lower bound which is the weight of the subgraph induced by the edges accepted in the node. In the bounding phase, we update the upper bound of the subgraph and we decide if this branch can lead to the optimum solution. This is done by comparing the upper bound with the current best lower bound. For the heaviest $k$-clique problem, we add the constraint that each subgraph that we keep has to be a clique, which is intuitive as every induced subgraph of the heaviest $k$-clique is also a clique. For the solution to contain the input query node $q$, we initialize each solution set with the query node $q$.

For the cocktail party problem, we implement the GREEDY algorithm presented in \cite{sozio10}. The algorithm can be summarized as follows: starting from an input graph $G$, at each step, the node of minimum degree is removed. If the node of minimum degree is $q$ or if the graph becomes disconnected, we stop. The solution is the the induced subgraph with the largest minimum degree. 
The cocktail party problem already allows for an input query node $q$ to be given in input, so no further modification is necessary. 
The algorithm for the cocktail party with distance $d$ works as follows: starting from an input graph $G$, at each step a node at distance bigger than $d$ from the query node $q$ is removed, or if no such node exists than the node of minimum degree is removed. We define the distance between two nodes to be the length of the shortest path and the degree to be the weighted degree. 


\subsection{Keyword coherence}

In order to evaluate which method produces the best set of keywords, we use a metric proposed for the automatic evaluation of the coherence of keywords describying a topic~\cite{Mimno2011}. The metric has been shown to be a good indicator of human judgment. 

Given a collection of tweets $C$, a set of keywords $K$, the frequencies of keywords $f(k_1)$ and keyword co-occurrences $f(k_1, k_2)$ in $C$, the coherence of $K$ given $C$ is:  
\begin{equation}
Coherence(K|C) = \sum_{k_1,k_2 \in K} \log(\frac{f(k_1, k_2) + 1}{f(k_1)})
\end{equation}

\subsection{Evaluation}
In order to assure a fair comparison, we run Algorithm~\ref{algo:events} while only changing line 10, that is the subroutine for computing keywords associated with the event. We considered the top 20 events in the months November 2015 - February 2016, in total 80 events. We compute the coherence of every event description given by the 6 methods (quasi clique, heaviest $k$-subgraph, heaviest $k$-clique, cocktail party problem, cocktail party problem with distance $d$ and top co-occurrences). 
For the heaviest $k$ subgraph and heaviest $k$-clique we choose $k = 10$. For the cocktail party with distance $d$ we experimented with $d=1$ and $d=2$ and we choose $d=1$ as it gave better results. For the top $k$ co-occurrences, we return the $9$ top co-occuring keywords with query node $d$. For each approach, the query node is the location and the co-occurrence graph is created from tweets written in the event timeframe. We compare these results with the results returned by running the quasi clique subroutine with $v = location$, $\gamma = 0.4, s= 10$. 

We report the results in Table~\ref{tab:kcoh}. A cell $(i,j)$ gives the percentage of events on which the method $i$ performed better or equally well in comparison with method $j$. The results show that our definition of a weighted quasi-clique ($WQC$) gives the best results in terms of keyword coherence. The second best technique is cocktail party with distance $1$, followed by top $k$ co-occurrences.

\begin{table}[!htbp]
\small
\centering
\begin{tabular}{|c|c|c|c|c|c|c|}
\hline
Method & $WQC$ & $HkS$ & $HkC$ & $CP$ & $CPD$  & $Topk$  \\
\hline
$WQC$ & - & 90\% & 86.25\% & 92.5\% & 68.75\% & 76.25\%\\
$HkS$ & 11.25\% & - & 18.75\% & 63.75\%& 16.25 & 13.75\\
$HkC$ & 17.5\% & 82.5\% & - & 78.75\%& 32.5\% & 32.5\%\\
$CP$  & 8.75\% & 37.5\%& 27.5\%& - &12.5\% & 11.25\\
$CPD$ & 40\% & 83.75\% & 70\%& 98.75\%& - & 57.5\% \\
$Topk$& 30\% & 87.5\%& 73.75\% & 90\% & 48.75\% & -\\
\hline
\end{tabular}
\\~\\~\\
\caption{Comparison of keyword coherence between $WQC, HkS, HkC, CP, CPD, Topk$. On each line, we have the percentage of events on which the method performed better or equally well in respect to the other approaches. For example, $WQC$ performed better or equally well on 90\% of event descriptions when compared with $HkS$.}
\label{tab:kcoh}
\end{table}

\section{Conclusions}\label{sec:future}
We presented \evidense, a graph-based approach for finding high-impact events in social media. In particular, we address the challenge of providing a succinct and informative description of the events retrieved with our approach. We focus on disaster events, while we discuss how our approach could be adapted to other classes of events.
Our extensive experimental evaluation over a period of $19$  months shows that our approach outperforms state-of-the-art approaches in terms of precision and fraction of unique events retrieved, while the description provided by our algorithm is succinct and contains the most relevant information such as the location, what happened and the timeframe. We also showed how to improve the results even further by incorporating results from mainstream media.
Given these results, we consider \evidense could represent a valuable tool for analyzing both social content and news articles from mainstream media, as well as for studying how they compare. It could also be used to boost the performance of other approaches. For example, the dataset we released could be used to create training sets for automatic classifiers.

\section{Acknowledgments}
Part of this work was done while Oana Balalau was a student at T\'el\'ecom ParisTech University. Carlos Castillo is partially funded by La Caixa project LCF/PR/PR16/11110009.
Mauro Sozio is partially funded by the French National Agency (ANR) under project FIELDS (ANR-15-CE23-0006).
\bigskip

\noindent\textbf{Data and code sharing.}
The ids of the tweets we use in our evaluation are publicly available, together with our code.\footnote{\url{https://github.com/nyxpho/evidense}}

\bibliographystyle{ACM-Reference-Format}
\bibliography{ref}

\end{document}